\documentclass[conference]{IEEEtran}
\IEEEoverridecommandlockouts
\usepackage{cite}
\usepackage{amsmath,amssymb,amsfonts}
\usepackage{graphicx}
\usepackage{textcomp}
\usepackage{xcolor}

\usepackage[hyphens]{url}
\usepackage[most]{tcolorbox}

\newtcolorbox{findingbox}{
    colback=gray!10,      
    colframe=gray!50,     
    arc=0mm,              
    boxrule=0.5pt,        
    left=5pt, right=5pt, top=5pt, bottom=5pt, 
    fontupper=\small,     
}

\def\BibTeX{{\rm B\kern-.05em{\sc i\kern-.025em b}\kern-.08em
    T\kern-.1667em\lower.7ex\hbox{E}\kern-.125emX}}
\begin{document}

\title{Reliability Characterization for N-version Object Detection\\}

\author{\IEEEauthorblockN{Shunsuke Nagao}
\IEEEauthorblockA{\textit{Department of Computer Science} \\
\textit{University of Tsukuba}\\
Tsukuba, Japan \\
nagao.shunsuke@sd.cs.tsukuba.ac.jp}
\and
\IEEEauthorblockN{Fumio Machida}
\IEEEauthorblockA{\textit{Department of Computer Science} \\
\textit{University of Tsukuba}\\
Tsukuba, Japan \\
machida@cs.tsukuba.ac.jp}
}

\maketitle
\thispagestyle{empty} 
\pagestyle{empty}     

\begin{abstract}
N-version object detection (OD) is an approach to diversifying detection results using multiple models or input frames and reducing detection errors by aggregating individual results.
Diversity and consistency across multiple detection results are critical information for characterizing the reliability of possible configurations of N-version OD systems.
However, existing performance metrics such as $mAP$ and $Accuracy$ fail to capture these factors, as they are defined solely on the final outcome after aggregation. 
To overcome this limitation, we propose two reliability metrics particularly defined for N-version OD, namely coverage of errors in OD ($Cov\_OD$) and certainty of accurate prediction in OD ($Cer\_OD$), which can be computed from individual detection results without relying on voting strategies. 
We empirically demonstrate the unique features of the proposed metrics through a case study of N-version OD for a vehicle in an autonomous-driving simulator. 
We show that the proposed metrics can be used for 
1) selecting version combinations with complementary error characteristics, 
2) choosing an effective voting strategy based on diversity and consistency profiles, and 
3) guiding incremental construction of N-version OD systems via pairwise two-version analysis. 
These results highlight the importance of the metrics that can guide the design of reliable N-version OD applications.
\end{abstract}

\begin{IEEEkeywords}
Autonomous driving, inference errors, N-version machine learning, object detection, reliability
\end{IEEEkeywords}

\section{Introduction}
\label{section-Introduction}
Object detection is used in many application fields, such as autonomous driving, surveillance cameras, and medical image diagnosis \cite{zou2023object, litjens2017survey, grigorescu2020survey}. 
Recent object detection systems are mostly based on machine learning (ML) models trained from labeled datasets. 
Since ML model predictions depend heavily on the quality of the training data and the algorithm, ML-based object detection is prone to various errors, such as false detections, misclassifications, and omissions.  
As detection errors can cause significant risk in safety-critical autonomous systems, ensuring the reliability of object detection is a critical challenge.

N-version ML system is an architectural approach to improve the reliability of ML system outputs using multiple versions of inference results~\cite{machida2019n} at runtime. 
Multiple ML-based classifiers and manipulated input image data were used to provide reliable image classification systems~\cite{machida2019diversity}. 
Multi-version inference approaches have also been adopted for malware detection~\cite{marques2021waste}, object detection~\cite{wakigami2024empirical}, and perception systems for autonomous vehicles~\cite{wen2024enhancing}.

Reliability improvements by N-version ML systems have been evaluated by several metrics.
Xu et al. introduced distinguished error rates (DER) and mutual error rates (MER) that measure the ratio of erroneous test cases occurring in only one model and the ratio of erroneous cases shared by all models, respectively~\cite{DSML2023Xu}. 
Machida~\cite{machida2019diversity} and Takahashi~\cite{prdc2022takahashi} used Coverage of Errors ($Cov$) and Certainty of Accurate Prediction ($Cer$) to quantify the error coverage and the consistency on correct inferences over N-version ML-based classifiers. 
Zoppi et al.~\cite{zoppi2023ensembling, prdc2024Zoppi} considered the omission probability for evaluating the safety of ensemble classifiers, where unreliable classification results are dropped for safe decisions.
These metrics primarily focus on classification tasks and are not directly applicable to evaluating the reliability of object detection, where many target objects can appear in a single image. 
For example, when many objects appear in a given frame but some are not properly detected, conventional metrics do not count multiple prediction errors individually. As a result, the inference on the frame is treated as an error, no matter how many objects are correctly detected.
On the other hand, conventional performance metrics for object detection tasks, such as $mAP$ and $F1\text{-}score$ are defined for a single object detector.
Since these metrics require a specific voting strategy to combine multiple inference results into a single output, their value depends on which voting strategy is applied, making it impossible to characterize the inherent reliability of an N-version configuration independently of the voting.
These gaps motivate us to consider finer-grained object-wise reliability metrics. 

In this study, we propose new metrics for characterizing the reliability of an N-version object detection system that enable finer-grained, object-wise reliability evaluation than conventional metrics. 
The proposed metrics, namely $Cov\_OD$ and $Cer\_OD$, quantify the ratio of detections covered by correct detections and the ratio of correct detections across all component versions in N-version inference results. 
$Cov$ stands for $Coverage$ and represents the degree to which the system's output is correct when an incorrect inference is compensated for by another correct inference.
$Cer$ stands for $Certainty$ and represents the degree to which an inference output by the system is certainly correct.
Unlike $Cov$ and $Cer$ presented for the classification task~\cite{machida2019diversity}, the proposed metrics quantify the intersections and unions of object-wise detection errors, enabling finer-grained reliability evaluation. 
Moreover, the proposed metrics can capture error overlap and the diversity of detection results without requiring a specific voting strategy, thereby overcoming the limitations of conventional performance metrics for a single object detector.
These unique features of the metrics enable an effective design of reliable N-version object detection systems.

To demonstrate the benefits of the proposed metrics, we conducted experiments using a dataset collected from an autonomous vehicle simulator Carla \cite{Carla}.  
We constructed diverse configurations of N-version object detection systems by using different object detection models or input image data obtained from multiple cameras mounted on a vehicle. 
The reliability of the generated diverse N-version object detection systems is evaluated using $Cov\_OD$ and $Cer\_OD$, as well as several conventional metrics for comparison.
The values of $Cov\_OD$ and $Cer\_OD$ are consistently shown to be higher than existing baseline counterparts $Cov$ and $Cer$, confirming that the proposed metrics can properly capture object-wise detection errors, allowing finer-grained evaluation.

To further demonstrate the benefits of $Cov\_OD$ and $Cer\_OD$ in constructing an effective N-version object detection system, we conducted exhaustive comparisons of possible configurations for two-version and three-version object detection systems using the same dataset. Through the detailed empirical comparative analysis, we show that the proposed metrics can be used for i) selecting effective version combinations with complementary error characteristics, ii) choosing an appropriate voting strategy based on diversity and consistency profiles, and iii) guiding the incremental construction of three-version object detection systems from pairwise two-version analysis. 

In short, the paper makes the following contributions.
\begin{itemize}
    \item We propose new metrics for characterizing the reliability of N-version object detection systems that quantify diversity and consistency across detection results on an object-wise basis, independently of any voting strategy.
    \item Through an empirical case study of an object detection system in an autonomous vehicle simulation, we demonstrate the advantages of the proposed metrics over conventional metrics.
    \item We derive practical guidelines for designing N-version object detection systems through the exhaustive comparative evaluations of two-version and three-version configurations.
\end{itemize}

The rest of the paper is organized as follows. 
Section \ref{section-Related work} describes related work. 
Section \ref{section-Existing metrics for Object Detection and limitations} reviews the existing evaluation metrics of object detection and clarifies the gap.
Section \ref{section-Proposed metrics} gives the definitions of the proposed metrics, $Cov\_OD$ and $Cer\_OD$. 
Section \ref{section-Evaluation} presents the experimental configuration and evaluation results. 
Section~\ref{section-design-insights} discusses design guides to compose two-version and three-version object detection systems.
Section~\ref{section-limitation} describes the limitations of the work.
Section \ref{section-conclusion} summarizes the paper and discusses future work.

\section{Related Works}
\label{section-Related work}
$mAP$ and $Accuracy$ are commonly used metrics for evaluating the performance of object detection tasks. 
As these basic metrics have several limitations when deployed in real systems, diverse metrics have been presented in the literature. 
A number of studies have proposed alternatives to $mAP$ by addressing its limitations from different perspectives.
Oksuz et al.\ \cite{LRP} proposed LRP, which jointly evaluates localization accuracy, recall, and precision.
Similarly, Ohtani et al.\ \cite{OC-Cost} introduced OC-Cost, which measures the corrective cost for classification and localization errors per image.
Other works focus on safety-oriented evaluations in autonomous driving. 
Ceccarelli et al.\ \cite{ceccarelli2023evaluating} proposed a metric incorporating object distance, collision possibility, and time-to-collision, while Ramanan et al.\ introduced R³ \cite{bansal2021risk}, which computes recall separately for different risk levels. 
Volk et al. \cite{comprehensiveSafetyMetrics} proposed a safety metric that integrates IoU based on distance, collision risk, and accident severity based on the RSS model, and time taken to recognize an object.
Gamerdinger et al. \cite{gamerdinger2025epsmnovelmetricevaluate} proposed a metric that takes into account the impact of both object detection safety and lane detection safety.
These metrics focus on the performance of one object detection model and are not directly applicable to N-version object detection that involves multiple detection results.

ML voting strategies are widely recognized as an effective approach for improving prediction accuracy by combining multiple ML models. The main ensemble strategies include bagging \cite{breiman1996bagging}, boosting \cite{freund1997decision,freund1996experiments}, and stacking \cite{wolpert1992stacked}, which reduce prediction errors by combining multiple models in different ways.
While these strategies are generally applicable to diverse ML tasks, additional consideration is required for object detection, where multiple target objects appear in a single input.

Voting strategies for object detection typically focus on combining multiple detection results into a single output.
Casado-Garcia et al.\ \cite{casado2020ensemble} introduced three decision rules for merging detections.
Other approaches improve bounding-box ensembling techniques. 
NMS removes overlapping boxes based on confidence, while PRAE \cite{mao2021probabilistic} addresses cross-model confidence inconsistencies by converting confidence scores into comparable probabilities.
Weighted Boxes Fusion (WBF) \cite{solovyev2021weighted} further retains and merges all bounding boxes via confidence-weighted averaging. 
These methods aim to improve the output by adopting a specific combination rule for multiple object detection results. 
In this study, we aim to evaluate the reliability and diversity of multiple object detection results without assuming a specific voting strategy.

\section{Limitations of existing metrics}
\label{section-Existing metrics for Object Detection and limitations}
In this section, we review existing metrics for object detection systems and discuss their limitations.

\subsection{Performance metrics for object detection task}
\label{subsection-Evaluation metrics for object detection model}
Conventional performance metrics such as $mAP$ and $Accuracy$ are commonly used to evaluate a single object detector. 
These metrics are calculated based on the object detection results, which are classified into four categories: True Positive (TP), False Positive (FP), False Negative (FN), and True Negative (TN).
$mAP$ is a metric that indicates how accurate an object detection model is at detection tasks. 
$Accuracy$ represents the ratio of correctly classified results to all predictions. 
Generally, $Accuracy$ is defined as $\frac{TP+TN}{TP+FP+FN+TN}$, but due to the difficulty of the definition of the number of $TN$, $Accuracy$ in the object detection task can be defined as $\frac{TP}{TP+FP+FN}$.

Since $Accuracy$ and $mAP$ are metrics to evaluate the performance of a single object detection model, for a system that involves multiple detectors, it is necessary to consider a voting strategy to determine the final output from the multiple inference results. 
Casado-Garcia et al. presented three voting strategies for combining inference results in an ensemble of object detectors \cite{casado2020ensemble}. 
The procedure is in three steps: 
At first, $N$ independent inference results are appended to a single list $F$. 
In the second step, the elements of $F$ are grouped into $G$ by the object class and the overlap of their bounding boxes. 
The overlap of bounding boxes is evaluated by $IoU$ for the given two bounding boxes $b1$ and $b2$.
\begin{equation}
    IoU(b1, b2) = \frac{area(b1 \cap b2)}{area(b1 \cup b2)}
\end{equation}
As a result, each element of $G$ is a list of inferences that have the same class and overlapping bounding boxes. 
In the final step, each element of $G$ is included in the final output based on either one of the following methods.
\begin{itemize}
    \item \textbf{affirmative method:} Include all elements of $G$ in the final output.
    \item \textbf{consensus method:} Include elements of $G$ whose number of elements is $\frac{N}{2}$ or more in the final output.
    \item \textbf{unanimous method:} Include the elements of $G$ with $N$ elements in the final output.
\end{itemize}

\begin{figure}[ht]
    \centering
    \includegraphics[width=0.9\linewidth]{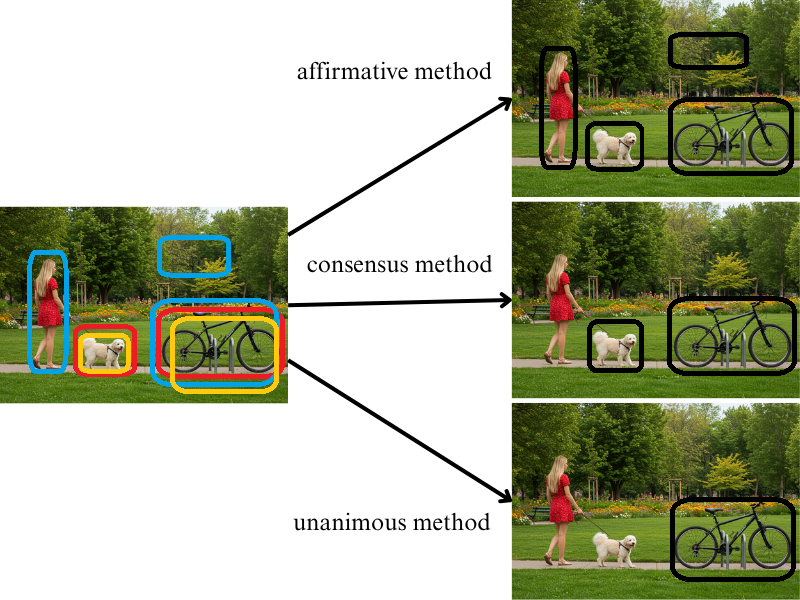}
    \caption{An example of three voting strategies for combining inference results in an ensemble of object detectors. 
    The bounding boxes in different colors represent the detection results by individual detectors.}
    \label{fig:example of voting strategy}
\end{figure}

Figure~\ref{fig:example of voting strategy} shows an example of three different voting strategies. 
By combining the inference results using any of the above methods, we can evaluate an N-version object detection system using $Accuracy$ and $mAP$. 
However, the affirmative method does not consider any conflicts between individual inference results, while the consensus and unanimity methods completely discard the influence of minority inference results that are not included in the final output. 
Therefore, the evaluation of $Accuracy$ and $mAP$ is tightly coupled with the adopted voting strategy, which may neglect inherent diversity gained by the N-version system.

\subsection{Reliability metrics for N-version ML systems}
To characterize the reliability of N-version ML systems, $Cov$ and $Cer$ have been proposed \cite{machida2019diversity}. 
$Cov$ is the percentage of input data for which one or more ML modules make correct inferences. 
Let $S$ be a set of input data, $M$ be a set of ML modules, each ML module be $m_i~(i = 1, 2, ...)$, and let $E_i$ be the input data set for which ML module $m_i$ makes an inference error on input data $S$. 
The absolute value symbol of the set represents the number of elements in the set. 
In this case, $Cov$ is given by the following formula.

\begin{equation}
    Cov = 1 - \frac{|\cap_{m_i \in M}E_i|}{|S|}
\end{equation}

On the other hand, $Cer$ is the percentage of input data for which all ML modules make correct inferences. In this case, $Cer$ is given by the following formula:

\begin{equation}
    Cer = 1 - \frac{|\cup_{m_i \in M}E_i|}{|S|}
\end{equation}

As the number of ML module versions increases, $Cov$ tends to improve because there are fewer errors common to all inference results. 
On the other hand, as the number of versions increases, $Cer$ tends to decrease because the probability that any inference result contains an incorrect inference increases. 
The reliability of an N-version ML system is characterized by these two metrics: 
$Cov$ indicates the probability that the correct output can be obtained by combining inferences, and $Cer$ indicates the probability that the entire system can reliably produce the correct output.

In contrast to accuracy metrics, $Cov$ and $Cer$ can incorporate the diversity of multiple inference results in the reliability evaluation. 
However, since $Cov$ and $Cer$ are metrics for image classification tasks, applying these metrics to object detection tasks may cause inappropriate evaluation. 
With $Cov$ and $Cer$, any inference errors in a single input image are counted as one inference error regardless of the number of detection errors, including false detections, misclassifications, and omissions.  

This means the total number of object detection errors is not accurately counted. 
Figure~\ref{fig:existing issue {multiple error}} shows an example. 
Suppose two inference results are obtained for an image containing three objects, represented by different color bounding boxes. 
One object (the person in the center) is included in both inference results, one object (the bike on the right) is included in one inference result but not in the other, and the other object (the dog in the center) is not included in either inference result. 
In this example, both $Cov$ and $Cer$ are equal to zero because no detection module makes a perfect detection for all target objects. 
This is a too rough reliability evaluation. 
Since two objects (the person and the bike) are included in more than one inference result and one object (the person) is included in two inference results, $Cov$ should be $\frac{2}{3}$ and $Cer$ should be $\frac{1}{3}$. 
We will consider such an object-wise fine-grained reliability evaluation by introducing new metrics.

\begin{figure}[ht]
    \centering
    \includegraphics[width=0.8\linewidth]{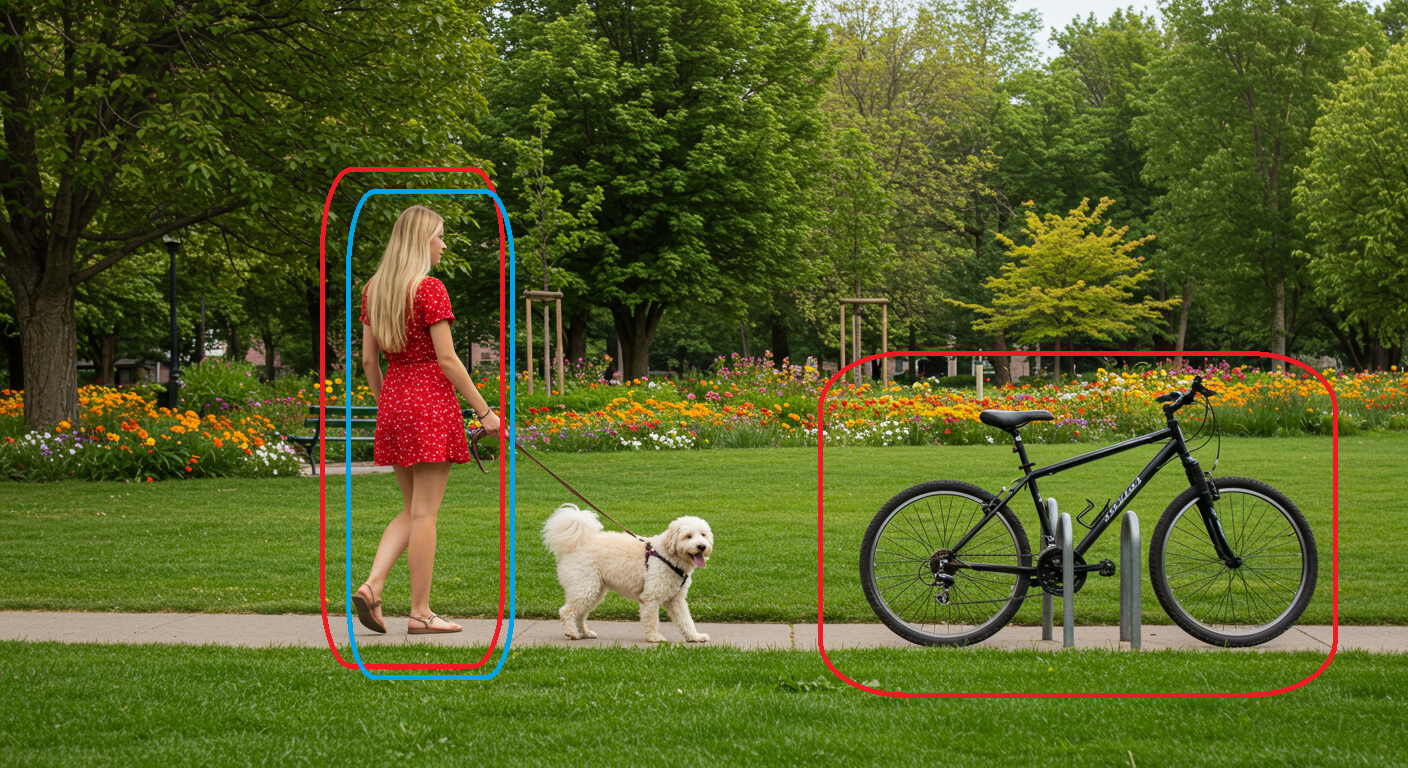}
    \caption{Existing metrics evaluate inference results on an image-by-image basis. Example: 
    One object (the person on the left) is detected by all detection modules, and one object (the bike on the right) is detected by only one module.}
    \label{fig:existing issue {multiple error}}
\end{figure}

\section{Proposed Metrics}
\label{section-Proposed metrics}
To overcome the limitation of the existing metrics, we propose new metrics for characterizing the reliability of N-version object detection systems.
Specifically, we extended the reliability metrics of N-version ML systems, $Cov$ and $Cer$, to consider the influence of multiple target objects in the input image.

Following the existing study \cite{machida2023using}, we consider the reliability of an N-version ML system as the probability that the system does not output errors. 
Reliability is defined in terms of the system output and is distinct from accuracy, which is a performance metric for the task. 
The system-level metric takes into account gray cases where the outputs are discarded or dropped to prevent potential errors that could cause system failure.

To define the new metrics, we first introduce the following notations. 
Let $M$ be the set of object detection modules in an N-version system, and $m_i~(i=1,2,...,N)$ be the elements of $M$. 
Define $TP_i(x)$, $FP_i(x)$, $FN_i(x)$ as the set of TP, FP, and FN generated by $m_i$ for input image $x$. 
The set of objects associated with the object detection by $m_i$ with input $x$ is given by the union.

\begin{equation}
    O_i(x) = TP_i(x) \cup FP_i(x) \cup FN_i(x)
\end{equation}

Let $O(x)$ be the union of $O_i(x)$ for all $m_i$.

\begin{equation}
    O(x) = \bigcup_{m_i \in M}O_i(x)
\end{equation}

To count the number of errors, we define $E_i(x)$ as the set of objects that are error detections output by $m_i$ for the input image $x$.

\begin{equation}
    E_i(x) = FP_i(x) \cup FN_i(x)
\end{equation}

Then, the intersection of errors $IoE(x)$ across all $m_i$ can be represented by

\begin{equation}
    IoE(x) = \bigcap_{m_i \in M}E_i(x)
\end{equation}

Similarly, the union of errors $UoE(x)$ across all $m_i$ is given by

\begin{equation}
    UoE(x) = \bigcup_{m_i \in M}E_i(x)
\end{equation}

Using the above definitions, we introduce the extended $Cov$ and $Cer$ for an N-version object detection system.

\subsection{Coverage of error in object detection}
We define the coverage of error in object detection as follows.
\begin{equation}
    \begin{aligned}
        Cov\_OD(M) &= 1 - \frac{1}{|S|}\sum_{x \in S}I(x), \\  
        I(x) &= 
            \begin{cases}
                \frac{|IoE(x)|}{|O(x)|} & \text{if } |O(x)| > 0, \\
                0 & \text{if } |O(x)| = 0.
            \end{cases}
    \end{aligned}
\end{equation}
    
$I(x)$ is the proportion of common incorrect inferences made by all object detection modules on the input image $x$. 
The higher $I(x)$ is, the greater the overlap in errors between the object detection modules.
$\frac{1}{|S|}\sum_{x \in S}I(x)$ is the average proportion of common incorrect inferences made by all object detection modules over the total input set. 
Therefore, $Cov\_OD$ represents the proportion of correct inferences made by at least one object detection module to the total size of the input image set. 
The higher $Cov\_OD$, the smaller the overlap in errors between each object detection module. 
$Cov\_OD$ represents the probability that the system will not produce an incorrect output by properly combining the inferences of the N-version object detection system.

\subsection{Certainty of accurate prediction in Object Detection}
The certainty of accurate prediction in object detection is defined
as follows.
\begin{equation}
    \begin{aligned}
        Cer\_OD(M) &= 1 - \frac{1}{|S|}\sum_{x \in S}U(x), \\
        U(x) &= 
        \begin{cases}
            \frac{|UoE(x)|}{|O(x)|} & \text{if } |O(x)| > 0,\\
            0 & \text{if } |O(x)| = 0.
        \end{cases}
    \end{aligned}
\end{equation}

$U(x)$ is the proportion of incorrect inferences made by one or more object detection modules on the input image $x$.
The smaller $U(x)$ is, the more commonly correct inferences each object detection model makes. $\frac{1}{|S|}\sum_{x \in S}U(x)$ is the average proportion of incorrect inferences in one or more $m_i$ across the entire input image.
Therefore, the $Cer\_OD$ represents the proportion of unanimity of correct inferences made by all object detection modules to the total size of the input image set. 
The higher $Cer\_OD$, the more the proportion of the unanimity of correct inferences is observed. 
$Cer\_OD$ represents the probability that the N-version object detection system will reliably produce no false outputs.

\subsection{Interpretation of Proposed Metrics}
$Cov\_OD$ can be interpreted as the accuracy assuming that all errors are covered to the greatest extent possible, that is, the maximum accuracy achievable with an N-version configuration. 
A value close to 1.0 indicates that accurate detection across the entire system is possible if an appropriate voting method is employed.
A higher value is therefore preferable, because it indicates that few errors are common to all modules.
$Cer\_OD$ can be interpreted as the accuracy assuming that all errors appear in the output, that is, the minimum accuracy achievable with an N-version configuration.
However, a higher value is not necessarily preferable.
A value close to 1.0 implies not only a low overall error rate and strong inter-module consensus but also a alignment in detection tendencies across modules, indicating that the redundancy benefits of the N-version configuration are limited.
It is therefore necessary to consider both metrics simultaneously.
An ideal N-version configuration exhibits high $Cov\_OD$ with sufficiently low $Cer\_OD$, indicating that high accuracy is achievable with minimal redundancy through appropriate voting.

The formal bounds on achievable reliability follow these interpretations.
For any voting strategy applied to system $M$, the achievable system reliability $Reliability(M)$ satisfies
\begin{equation}
    Cer\_OD(M) \leq Reliability(M) \leq Cov\_OD(M).
    \label{eq:bounds}
\end{equation}
The upper bound follows because common errors cannot be recovered by any voting strategy, whereas the lower bound follows because unanimously correct detections are never discarded.
The gap $Cov\_OD(M) - Cer\_OD(M)$ defines the range of reliability exploitable by the choice of voting strategy, and reflects the degree of complementarity among modules.

\subsection{Metrics Comparison}
\begin{figure}[ht]
    \centering
    \includegraphics[width=\linewidth]{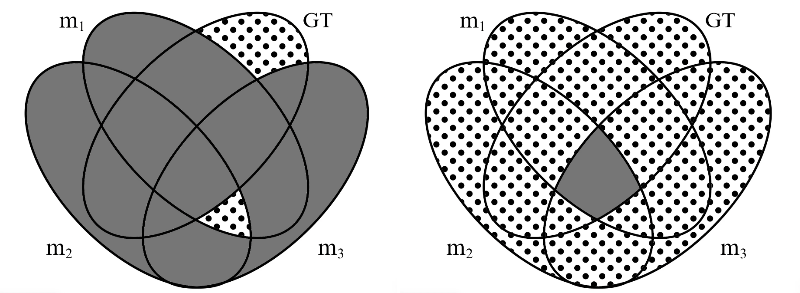}
    \caption{
    Venn diagram representation of Cov\_OD (left) and Cer\_OD (right) for a three-version object detection. 
    Cov\_OD and Cer\_OD correspond to the ratio of the grayed region over the grayed and dotted region.
    }
    \label{fig:venn ODmetrics}
    
    \centering
    \includegraphics[width=\linewidth]{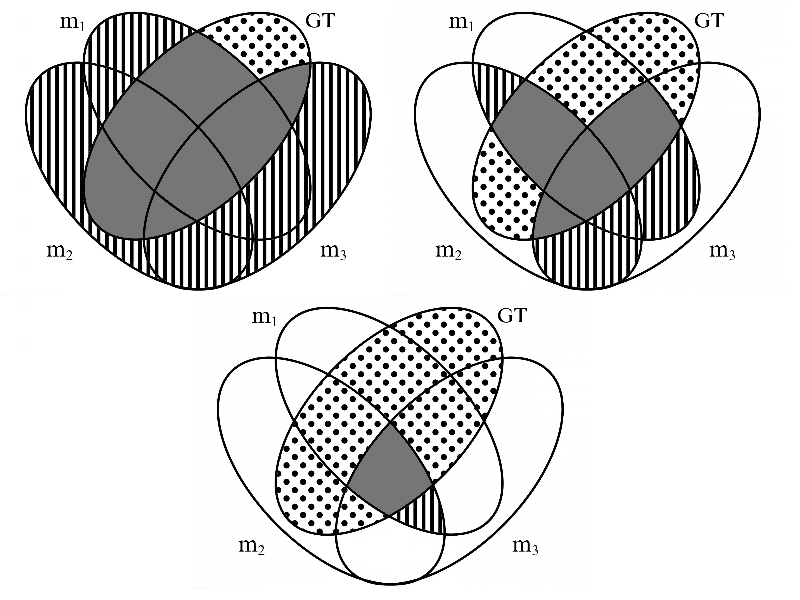}
    \caption{
    Venn diagram representation of performance metrics for a three-version object detection. The diagrams show the affirmative method (upper left), consensus method (upper right), and unanimous method (bottom center). The grayed, striped, and dotted regions correspond to TP, FP, and FN in the combined inference of each method, respectively. Object detection performance metrics are calculated based on each TP, FP, and FN information.
    }
    \label{fig:venn Ensemble}
\end{figure}

To illustrate the unique features of the proposed metric, we visualize the coverage relationship of the detection results in a three-version object detection system using Venn diagrams in Figures~\ref{fig:venn ODmetrics} and~\ref{fig:venn Ensemble}. 
$GT$ (Ground Truth) represents the set of objects present in the images, and $m_i~(i=1,2,3)$ denotes the set of objects detected by the object detection module $i$. 
The region of $m_i \cap GT$ represents the set of TP for $m_i$, $m_i \cap \overline{GT}$ is the set of FP, and $\overline{m_i} \cap GT$ is the set of FN. 

Figure~\ref{fig:venn ODmetrics} illustrates the Venn diagrams corresponding to $Cov\_OD$ and $Cer\_OD$. 
The grayed region in $Cov\_OD$ represents the set of inferences that can logically yield a correct system output, since at least one module provides a correct inference. 
This includes objects correctly detected by a single module, and module-specific FPs that do not appear in other modules and thus can be suppressed when combining multiple inferences. 
In contrast, the grayed region in $Cer\_OD$ represents the set of inferences that logically guarantee a correct system output, since all modules unanimously provide a correct detection. 
The dotted region in $Cov\_OD$ corresponds to $IoE(x)$ in (7), denoting FP or FN that are common to all modules, while the dotted region in $Cer\_OD$ corresponds to $UoE(x)$ in (8), denoting the union of all FP and FN across modules. 
The proposed metrics shown in Figure~\ref{fig:venn ODmetrics} compute the proportion of grayed regions relative to the $GT \cup m_1 \cup m_2 \cup m_3$.

On the other hand, Figure~\ref{fig:venn Ensemble} illustrates the Venn diagrams corresponding to the TP, FP, and FN obtained from the combined inference results using three voting strategies: affirmative, consensus, and unanimous methods.
In the figure, the grayed, striped, and dotted regions correspond to TP, 
FP, and FN, respectively, while the uncolored regions represent discarded inferences. Performance metrics for object detection tasks, such as $mAP$, are calculated based on the TP, FP, and FN information obtained in this way.
Consequently, the evaluation depends on the specific voting strategy employed, failing to reflect the diversity of errors and correct inferences across the modules.

$Cov\_OD$ characterizes system reliability by measuring how often at least one module correctly detects each object.
This captures the idea that a failure in one module (FN or FP) can be compensated by correct detections from other modules.
In contrast, $Cer\_OD$ characterizes system reliability by measuring how often all modules simultaneously detect the same object correctly.
This captures the idea that a correct detection in one module (TP) can be covered by false detections from other modules.
Thereby, $Cov\_OD$ and $Cer\_OD$ explicitly incorporate coverage relationships between correct and incorrect inferences. 
Furthermore, they incorporate the
diversity of inferences into the evaluation, as they reflect the inferences of even a single module or a minority of modules.

The comparison shows that the proposed metrics consider the diversity of inferences by the N-version configuration and the covering relationship of errors, which could not be captured by conventional metrics.

\section{Evaluation}
\label{section-Evaluation}
In this section, we demonstrate the benefits of the proposed $Cov\_OD$ and $Cer\_OD$ through a case study of an N-version object detection system in an autonomous driving scenario. 
The evaluation is conducted with the original image dataset collected from the Carla simulator.

\subsection{Dataset and Construction of N-version Object Detection Systems}
Carla is an open-source simulator for autonomous driving systems~\cite{Carla}. 
The simulator can create various autonomous driving environments by deploying cars and pedestrians in different weather conditions and at different times of day. 
In this study, we used Carla 0.9.13 Release and conducted experiments in Town03. 
CARLA automatically generated ground-truth bounding boxes for all objects. This dataset serves as the common basis for constructing N-version object detection systems with different diversity sources.

\begin{figure}[ht]
    \centering
    \includegraphics[width=\linewidth]{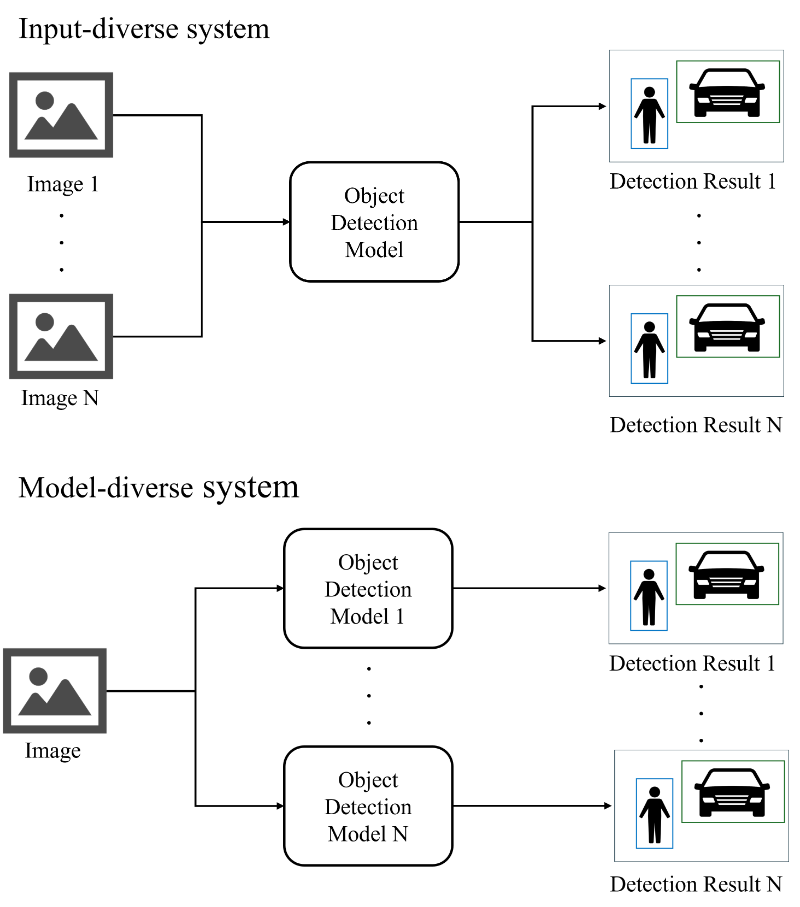}
    \caption{
    Two types of N-version object detection system configurations used in the evaluation.
    (top)~The input-diverse system feeds $N$ images into a single detection model.
    (bottom)~The model-diverse system feeds a single image into $N$ independent detection models.}
    \label{fig:nversion setting}
\end{figure}

\subsubsection{Input-diverse N-version system}
To collect diverse input images, we installed five RGB cameras in front of the ego vehicle in the simulator environment. 
Table~\ref{table: dataset camera location} shows the positions of the five cameras. 
The coordinates of the camera positions are expressed by $(x, y, z)$, where $x$ is the longitudinal position from the vehicle's bottom center, $y$ is the lateral position, and $z$ is the height. 
The viewing angle of all cameras was set to 60 degrees. Each camera captured 2000 images during the driving scenario. 
Object detection was performed using YOLOv8n~\cite{yolov8} for all camera inputs.
The N-version object detection system was constructed by incrementally adding camera views in the IDs of the cameras in Table 1.
This setting allows us to analyze how the proposed metrics respond when the same model receives visually different inputs.

\begin{table}[ht]
    \caption{camera location}
    \centering
    \begin{tabular}{lll} 
        \hline
        Camera ID & Camera name & Location(x,y,z)(m) \\ \hline
        1 & center & (1.5, 0.0, 2.0) \\
        2 & left1 & (1.5, -0.3, 2.0) \\
        3 & right1 & (1.5, 0.3, 2.0) \\
        4 & left2 & (1.5, -0.6,2.0) \\
        5 & right2  & (1.5, 0.6, 2.0) \\
        \hline
    \end{tabular}
    \label{table: dataset camera location}
\end{table}

\subsubsection{Model-diverse N-version system}
We also constructed an N-version object detection system based on different object detection models. 
In addition to YOLOv8n, RT-DETR~\cite{rtdetr}, RetinaNet with ResNet-50 FPN backbone~\cite{retinanet}, FCOS with ResNet-50 FPN backbone~\cite{fcos}, and SSD with VGG-16 backbone~\cite{ssd} were used to obtain diverse predictions and added to the system in this order. 
The input image used for all models was 1.center. 
This setting enables us to evaluate how the metrics behave when multiple models produce varied detection outputs for the same input.

The dataset, including ground-truth annotations and all detection results for both diversity settings, is publicly available for reproducibility\footnote{\url{https://anonymous.4open.science/r/Reliability-Characterization-for-Nversion-Object-Detection-04D2/}}.

\subsection{Reliability metrics}
We evaluate the reliability of different versions of N-version object detection systems by $Cov$, $Cer$, $Cov\_OD$, and $Cer\_OD$. 
We also compare the evaluation results of $Cov\_OD$ and $Cer\_OD$ with $mAP$ using three different voting strategies discussed in Section \ref{subsection-Evaluation metrics for object detection model}. 
We use $IoU$ to determine the overlap of bounding boxes, with a threshold of 0.5.

\subsection{Single-Version Baseline Evaluation}
\label{subsec:baseline}

\begin{table}[ht]
    \centering
    \caption{Single-version baseline $mAP$ for each model and YOLOv8n with each camera view}
    \label{tab:baseline}
    \begin{minipage}[t]{0.46\linewidth}
        \centering
        \small
        \textit{(a) Model baseline}\\[4pt]
        \begin{tabular}{lc}
            \hline
            Model & $mAP$ \\
            \hline
            YOLOv8n   & 0.522 \\
            RT-DETR & 0.530 \\
            RetinaNet & 0.588 \\
            FCOS      & 0.570 \\
            SSD       & 0.427 \\
            \hline
        \end{tabular}
    \end{minipage}
    \hfill
    \begin{minipage}[t]{0.46\linewidth}
        \centering
        \small
        \textit{(b) Each View baseline}\\[4pt]
        \begin{tabular}{lc}
            \hline
            Camera view & $mAP$ \\
            \hline
            Center  & 0.522 \\
            left1  & 0.394 \\
            right1 & 0.385 \\
            left2  & 0.206 \\
            right2 & 0.243 \\
            \hline
        \end{tabular}
    \end{minipage}
\end{table}

Table~\ref{tab:baseline} reports the single-version $mAP$ for each model and camera configuration, which serves as the baseline reliability for N-version object detection systems.
Table~\ref{tab:baseline} (a) shows the results for each model's detection performance, where each of the five models processes the same center camera input.
RetinaNet achieves the highest $mAP$ of 0.588, followed by FCOS (0.570) and RT-DETR (0.530).
YOLOv8n shows moderate performance (0.522), while SSD (0.427) yields the lowest value.

Table~\ref{tab:baseline} (b) shows the performance of YOLOv8n with different camera views. 
For a consistent comparison, the evaluation is conducted based on the objects visible in the center camera view as the ground truth.
While the case using the center camera achieves $mAP$ of 0.522, configurations using horizontally offset cameras show a performance decline as the displacement from the center increases.
left1 and right1 achieve 0.394 and 0.385, and the further-offset left2 and right2 drop to 0.206 and 0.243.
This degradation occurs because objects captured by the center camera may be partially displaced or occluded in offset views, reducing per-camera detection accuracy.
Despite this per-camera accuracy loss, the offset views capture complementary spatial information, which motivates their use in input-diverse N-version object detection systems to compensate for each other's detection errors.

\subsection{Comparison with $Cov$ and $Cer$}
\begin{table}[t]
    \centering
    \caption{Comparison of $Cov$, $Cov\_OD$, $Cer$, and $Cer\_OD$
             for the input-diverse N-version system}
    \label{tab:Cov Cer comparison input}
    \begin{tabular}{lcccc}
        \hline
        Num of inputs & $Cov$ & $Cov\_OD$ & $Cer$ & $Cer\_OD$ \\
        \hline
        1 & 0.689 & 0.806 & 0.689 & 0.806 \\
        2 & 0.713 & 0.845 & 0.655 & 0.767 \\
        3 & 0.727 & 0.867 & 0.629 & 0.739 \\
        4 & 0.735 & 0.886 & 0.609 & 0.704 \\
        5 & 0.742 & 0.903 & 0.599 & 0.687 \\
        \hline
    \end{tabular}
\end{table}

\begin{table}[t]
    \centering
    \caption{Comparison of $Cov$, $Cov\_OD$, $Cer$, and $Cer\_OD$
             for the model-diverse N-version system}
    \label{tab:Cov Cer comparison model}
    \begin{tabular}{lcccc}
        \hline
        Num of models & $Cov$ & $Cov\_OD$ & $Cer$ & $Cer\_OD$ \\
        \hline
        1 & 0.689 & 0.806 & 0.689 & 0.806 \\
        2 & 0.710 & 0.904 & 0.584 & 0.725 \\
        3 & 0.751 & 0.932 & 0.572 & 0.700 \\
        4 & 0.764 & 0.950 & 0.550 & 0.667 \\
        5 & 0.771 & 0.956 & 0.532 & 0.628 \\
        \hline
    \end{tabular}
\end{table}

Table~\ref{tab:Cov Cer comparison input} shows the comparison results between the existing reliability metrics ($Cov$ and $Cer$) and the proposed metrics ($Cov\_OD$ and $Cer\_OD$) for input-diverse N-version object detection systems.
In the input-diverse systems, the number of versions increases from 1 to 5 by sequentially adding camera inputs in order of their IDs described in Table~\ref{table: dataset camera location}. 
As the number of versions increases from 1 to 5, both $Cov$ and $Cov\_OD$ increase, while $Cer$ and $Cer\_OD$ decrease.
This reflects the fundamental trade-off of N-version object detection systems: obtaining a wider variety of detection results can cover more errors, while simultaneously reducing the likelihood of shared accurate detections.

Compared with $Cov$, $Cov\_OD$ consistently yields higher values across all versions.
The results indicate that $Cov\_OD$ can evaluate the error coverage more precisely by counting detection errors in individual objects rather than frames. $Cov$ fails to count the partial success cases where some objects in the frame are not correctly detected, causing the underestimate of error coverage.
Similarly, we can observe that $Cer\_OD$ is consistently higher than the corresponding $Cer$, showing the advantage of $Cer\_OD$ in a finer-grained evaluation. $Cer$ tends to underestimate the consistency of correct detections because it does not count the partially consistent cases where only some objects in the image are consistently detected by all modules.

Table~\ref{tab:Cov Cer comparison model} shows the results for the model-diverse N-version object detection systems. 
In this case, the number of versions increases by adding distinct object detection models while using the same input image.
We observed the same trend as with input-diverse systems.
As the number of versions increases from 1 to 5, $Cov\_OD$ remains consistently higher than $Cov$, and $Cer\_OD$ remains consistently higher than $Cer$, for all numbers of versions.
These results confirm that the advantages of the proposed metrics over the existing $Cov$ and $Cer$, namely, finer-grained object-level evaluation, are not specific to the input-diverse setting but also hold for model-diverse systems.

\begin{findingbox}
    \textbf{Findings 1:}
    $Cov\_OD$ and $Cer\_OD$ enable finer-grained reliability assessment than $Cov$ and $Cer$ by evaluating detection correctness at the object level, preventing underestimation caused by a small number of consistently undetectable objects dominating the image-level evaluation.
\end{findingbox}

\subsection{Comparison with mAP}
\begin{table}[t]
    \centering
    \caption{Comparison of $Cov\_OD$, $Cer\_OD$, and $mAP$ under three
             voting strategies for the input-diverse N-version system.}
    \label{tab:mAP comparison input}
    \begin{tabular}{lccccc}
        \hline
        & & & \multicolumn{3}{c}{$mAP$} \\
        \cline{4-6}
        Num of inputs & $Cov\_OD$ & $Cer\_OD$ & Affir. & Cons. & Unanim. \\
        \hline
        1 & 0.806 & 0.806 & 0.522 & 0.522 & 0.522 \\
        2 & 0.845 & 0.765 & 0.436 & 0.436 & 0.332 \\
        3 & 0.867 & 0.739 & 0.434 & 0.402 & 0.331 \\
        4 & 0.886 & 0.704 & 0.406 & 0.394 & 0.308 \\
        5 & 0.903 & 0.687 & 0.387 & 0.374 & 0.308 \\
        \hline
    \end{tabular}
\end{table}

\begin{table}[t]
    \centering
    \caption{Comparison of $Cov\_OD$, $Cer\_OD$, and $mAP$ under three
             voting strategies for the model-diverse N-version system}
    \label{tab:mAP comparison model}
    \begin{tabular}{lccccc}
        \hline
        & & & \multicolumn{3}{c}{$mAP$} \\
        \cline{4-6}
        Num of models & $Cov\_OD$ & $Cer\_OD$ & Affir. & Cons. & Unanim. \\
        \hline
        1 & 0.806 & 0.806 & 0.522 & 0.522 & 0.522 \\
        2 & 0.904 & 0.725 & 0.580 & 0.580 & 0.521 \\
        3 & 0.932 & 0.700 & 0.584 & 0.560 & 0.481 \\
        4 & 0.950 & 0.667 & 0.588 & 0.609 & 0.493 \\
        5 & 0.956 & 0.628 & 0.563 & 0.559 & 0.410 \\
        \hline
    \end{tabular}
\end{table}
The conventional performance metrics for object detection are fundamentally limited for evaluating N-version systems, as their value is inseparable from the chosen voting strategy. 
To demonstrate the voting-agnostic property of the proposed metrics, we compare the proposed metrics with $mAP$ computed under different voting strategies (affirmative, consensus, unanimous) for the input-diverse N-version object detection systems.
In Table V, as the number of versions increases from 1 to 5, $mAP$ under each voting strategy decreases monotonically, similar to the trend of $Cer\_OD$.
However, the values of $mAP$ and the degree of degradation depend on the voting strategy. 
$mAP$ under unanimous voting is more significantly affected by the number of versions, compared with the cases with affirmative and consensus methods.

Table~\ref{tab:mAP comparison model} shows the results for the model-diverse N-version object detection systems.
In contrast to the input-diverse case, $mAP$ does not show a consistent trend across voting strategies.
Under the affirmative and consensus methods, $mAP$ improves as the number of versions increases up to four versions, while it decreases at five versions.
Despite these distinctive behaviors across voting strategies, the proposed metrics exhibit monotonic trends, regardless of the voting strategy.

The results indicate that $mAP$ does not consistently capture the diversity and consistency characteristics of N-version object detection systems, as its value is significantly affected by the choice of voting strategy and varies with the diversity setting.
This highlights a fundamental limitation of $mAP$: it is difficult to disentangle whether performance changes are due to the N-version configuration itself or to the choice of voting strategy.
In particular, even when detection outputs from different inputs or models successfully detect different subsets of objects, certain voting strategies may discard these complementary detections, resulting in an underestimation of $mAP$.
In contrast, $Cov\_OD$ and $Cer\_OD$ provide a consistent, voting-agnostic characterization of the reliability of N-version object detection systems.
$Cov\_OD$ confirms that an increasing number of versions improves system-level error coverage, while $Cer\_OD$ reveals that achieving consistent correct detections across all detections becomes increasingly difficult as diversity grows.

\begin{findingbox}
    \textbf{Findings 2:} 
    Unlike $mAP$, $Cov\_OD$ and $Cer\_OD$ can indicate the range of achievable reliability without assuming any specific voting strategy. 
    The proposed metrics characterize the inherent diversity and consistency of an N-version configuration.
\end{findingbox}


\section{Practical Design Insights from Proposed Metrics} 
\label{section-design-insights}

In this section, we analyze two- and three-version systems to understand how diversity and consistency among models affect system reliability. 
Based on the proposed metrics, we further derive practical design insights for selecting model combinations, choosing voting strategies, and extending configurations to more effective N-version object detection systems.
The goal of this section is to derive actionable design insights that guide practitioners in 
(i)~selecting model combinations, 
(ii)~choosing an appropriate voting strategy, and 
(iii)~incrementally extending a configuration toward more effective N-version object detection systems.

\subsection{Experimental Setup}
We used the same dataset as in the previous section.
In the preceding section, we evaluated the proposed metrics using a fixed pool of five detection models.
To provide a more comprehensive basis for design insights, we extend the analysis by adding two models, YOLOv5n~\cite{yolov5} and YOLOv11n~\cite{yolov11}, to create a pool of seven candidates.

All $_7C_2 = 21$ two-version and $_7C_3 = 35$ three-version object detection systems were evaluated.
For each configuration, we computed $Cov\_OD$, $Cer\_OD$, and $mAP$ under the affirmative, consensus, and unanimous methods to determine the final outputs, following the procedure described in Section~\ref{section-Evaluation}.

\subsection{Results for Two and Three-Version Systems}
\begin{table*}[ht]
    \centering
    \small
    \caption{Performance metrics for all 21 two-version object detection systems}
    \label{tab:two_version_od_map_comparison}
    \begin{tabular}{lccccc}
        \hline
        Combination & Cov\_OD & Cer\_OD & mAP\_affir. & mAP\_con. & mAP\_unanim. \\
        \hline
        YOLOv8n, RT-DETR     & 0.9041          & 0.7255          & 0.5810          & - & 0.5212 \\
        YOLOv8n, RetinaNet   & 0.8997          & \textbf{0.7614} & \textbf{0.5891} & - & 0.4833 \\
        YOLOv8n, FCOS        & \textbf{0.9221} & 0.7290          & 0.5847          & - & 0.5220 \\
        YOLOv8n, SSD         & 0.8468          & 0.7248          & 0.4865          & - & 0.4387 \\
        YOLOv8n, YOLOv5n     & 0.8445          & 0.7499          & 0.4957          & - & 0.4615 \\
        YOLOv8n, YOLOv11n    & 0.8548          & \textbf{0.7683} & 0.5085          & - & 0.4716 \\
        RT-DETR, RetinaNet   & 0.9183          & 0.7313          & \textbf{0.5910} & - & 0.5286 \\
        RT-DETR, FCOS        & 0.9171          & 0.7123          & 0.5737          & - & \textbf{0.5545} \\
        RT-DETR, SSD         & 0.9100          & 0.6751          & 0.5499          & - & 0.4468 \\
        RT-DETR, YOLOv5n     & 0.9092          & 0.6995          & 0.5423          & - & 0.4361 \\
        RT-DETR, YOLOv11n    & 0.9052          & 0.7193          & 0.5515          & - & 0.5069 \\
        RetinaNet, FCOS      & 0.9133          & 0.7492          & 0.5878          & - & \textbf{0.5694} \\
        RetinaNet, SSD       & 0.8906          & 0.7111          & 0.5748          & - & 0.4529 \\
        RetinaNet, YOLOv5n   & 0.8796          & 0.7473          & 0.5605          & - & 0.4768 \\
        RetinaNet, YOLOv11n  & 0.9006          & \textbf{0.7587} & \textbf{0.5910} & - & \textbf{0.5317} \\
        FCOS, SSD            & \textbf{0.9294} & 0.6780          & 0.5512          & - & 0.4449 \\
        FCOS, YOLOv5n        & 0.9200          & 0.7068          & 0.5717          & - & 0.4477 \\
        FCOS, YOLOv11n       & \textbf{0.9284} & 0.7213          & 0.5764          & - & 0.5155 \\
        SSD, YOLOv5n         & 0.8417          & 0.7017          & 0.4733          & - & 0.3962 \\
        SSD, YOLOv11n        & 0.8580          & 0.7153          & 0.4775          & - & 0.4426 \\
        YOLOv5n, YOLOv11n    & 0.8490          & 0.7447          & 0.4850          & - & 0.4540 \\
        \hline
    \end{tabular}
\end{table*}

\begin{table*}[ht]
    \centering
    \small
    \caption{Performance metrics for all three-version object detection systems}
    \label{tab:three_version_od_map_comparison}
    \begin{tabular}{lccccc}
        \hline
        Combination & Cov\_OD & Cer\_OD & mAP\_affir. & mAP\_con. & mAP\_unanim. \\
        \hline
        YOLOv8n, RT-DETR, RetinaNet   & 0.9321 & 0.7004 & 0.5844 & \textbf{0.5609} & 0.4815 \\
        YOLOv8n, RT-DETR, FCOS        & \textbf{0.9431} & 0.6829 & 0.5779 & 0.5607 & 0.4840 \\
        YOLOv8n, RT-DETR, SSD         & 0.9231 & 0.6690 & 0.5505 & 0.5028 & 0.4393 \\
        YOLOv8n, RT-DETR, YOLOv5n     & 0.9205 & 0.6914 & 0.5552 & 0.5024 & 0.4392 \\
        YOLOv8n, RT-DETR, YOLOv11n    & 0.9200 & 0.7065 & 0.5596 & 0.5053 & 0.4680 \\
        YOLOv8n, RetinaNet, FCOS      & 0.9386 & 0.7068 & \textbf{0.5928} & 0.5707 & \textbf{0.4932} \\
        YOLOv8n, RetinaNet, SSD       & 0.9109 & 0.6970 & 0.5793 & 0.5240 & 0.4135 \\
        YOLOv8n, RetinaNet, YOLOv5n   & 0.9068 & \textbf{0.7289} & 0.5836 & 0.5183 & 0.4803 \\
        YOLOv8n, RetinaNet, YOLOv11n  & 0.9136 & \textbf{0.7369} & 0.5851 & 0.5197 & 0.4847 \\
        YOLOv8n, FCOS, SSD            & 0.9337 & 0.6677 & 0.5545 & 0.5043 & 0.4443 \\
        YOLOv8n, FCOS, YOLOv5n        & 0.9291 & 0.6960 & 0.5826 & 0.5227 & 0.4597 \\
        YOLOv8n, FCOS, YOLOv11n       & 0.9319 & 0.7054 & 0.5784 & 0.5155 & 0.4794 \\
        YOLOv8n, SSD, YOLOv5n         & 0.8630 & 0.6926 & 0.4893 & 0.5063 & 0.4023 \\
        YOLOv8n, SSD, YOLOv11n        & 0.8752 & 0.7067 & 0.5219 & 0.5002 & 0.4380 \\
        YOLOv8n, YOLOv5n, YOLOv11n    & 0.8690 & \textbf{0.7327} & 0.4870 & 0.4938 & 0.4311 \\
        RT-DETR, RetinaNet, FCOS      & \textbf{0.9418} & 0.6886 & \textbf{0.5873} & \textbf{0.5714} & \textbf{0.5330} \\
        RT-DETR, RetinaNet, SSD       & 0.9343 & 0.6577 & 0.5746 & \textbf{0.5633} & 0.4123 \\
        RT-DETR, RetinaNet, YOLOv5n   & 0.9309 & 0.6864 & 0.5783 & 0.5553 & 0.4770 \\
        RT-DETR, RetinaNet, YOLOv11n  & 0.9350 & 0.6964 & 0.5790 & 0.5579 & 0.4818 \\
        RT-DETR, FCOS, SSD            & \textbf{0.9465} & 0.6412 & 0.5559 & 0.5572 & 0.4117 \\
        RT-DETR, FCOS, YOLOv5n        & 0.9407 & 0.6651 & 0.5702 & 0.5562 & 0.4533 \\
        RT-DETR, FCOS, YOLOv11n       & 0.9420 & 0.6768 & 0.5692 & 0.5602 & 0.4821 \\
        RT-DETR, SSD, YOLOv5n         & 0.9227 & 0.6511 & 0.5340 & 0.5010 & 0.3966 \\
        RT-DETR, SSD, YOLOv11n        & 0.9246 & 0.6630 & 0.5334 & 0.5001 & 0.3933 \\
        RT-DETR, YOLOv5n, YOLOv11n    & 0.9196 & 0.6852 & 0.5426 & 0.4981 & 0.4604 \\
        RetinaNet, FCOS, SSD          & 0.9401 & 0.6667 & 0.5618 & 0.5556 & 0.4560 \\
        RetinaNet, FCOS, YOLOv5n      & 0.9312 & 0.6983 & 0.5846 & 0.5641 & 0.4648 \\
        RetinaNet, FCOS, YOLOv11n     & 0.9397 & 0.7057 & \textbf{0.5858} & 0.5679 & \textbf{0.4924} \\
        RetinaNet, SSD, YOLOv5n       & 0.9027 & 0.6872 & 0.5724 & 0.4804 & 0.4139 \\
        RetinaNet, SSD, YOLOv11n      & 0.9120 & 0.6917 & 0.5703 & 0.5233 & 0.4142 \\
        RetinaNet, YOLOv5n, YOLOv11n  & 0.9087 & 0.7267 & 0.5720 & 0.5062 & 0.4705 \\
        FCOS, SSD, YOLOv5n            & 0.9370 & 0.6541 & 0.5481 & 0.5026 & 0.3987 \\
        FCOS, SSD, YOLOv11n           & 0.9407 & 0.6611 & 0.5467 & 0.5095 & 0.4459 \\
        FCOS, YOLOv5n, YOLOv11n       & 0.9348 & 0.6886 & 0.5723 & 0.5140 & 0.4494 \\
        SSD, YOLOv5n, YOLOv11n        & 0.8742 & 0.6891 & 0.4791 & 0.5021 & 0.4014 \\
        \hline
    \end{tabular}
\end{table*}

Tables~\ref{tab:two_version_od_map_comparison} and~\ref{tab:three_version_od_map_comparison} summarize the results of all two-version and three-version object detection systems, respectively.

In both cases, combinations involving architecturally heterogeneous models achieve the highest $Cov\_OD$ values.
For example, (FCOS, SSD) and (FCOS, YOLOv11n) reach $Cov\_OD$ of 0.9294 and 0.9284 among two-version systems, while (RT-DETR, FCOS, SSD) and (YOLOv8n, RT-DETR, FCOS) exhibit the highest $Cov\_OD$ values (0.9465 and 0.9431) among three-version systems.
In contrast, combinations dominated by similar architectures, particularly within the YOLO family, consistently yield lower $Cov\_OD$ values, indicating that increasing the number of similar models does not substantially improve error diversity, as their error patterns remain correlated.

For $Cer\_OD$, combinations of structurally similar models, especially within the YOLO family, tend to achieve the highest values.
For instance, (YOLOv8n, YOLOv11n) among two-version systems and (YOLOv8n, RetinaNet, YOLOv11n) among three-version systems achieve relatively high $Cer\_OD$, reflecting stronger agreement in correct detections.
Conversely, combinations such as (RT-DETR, SSD) and (FCOS, SSD) result in lower $Cer\_OD$ due to architectural differences.
Notably, $Cer\_OD$ generally decreases when extending from two-version to three-version systems, since the probability that all models simultaneously produce correct detections decreases as architectural diversity increases.

We further observe that combinations with relatively low $Cov\_OD$ and high $Cer\_OD$ tend to show little variation in $mAP$ across voting strategies, and often yield consistently low $mAP$ overall.
This indicates that such configurations have limited complementary error coverage due to a lack of diversity
Consequently, the choice of voting strategy has little effect and the benefit of multi-version integration remains limited.
This observation highlights a limitation of relying solely on $mAP$, as stable or low $mAP$ values do not by themselves reveal whether a configuration lacks diversity or error coverage.
Therefore, when designing N-version object detection systems, it is important to verify sufficient diversity and error coverage using $Cov\_OD$ and $Cer\_OD$, rather than relying on $mAP$ alone.

\begin{findingbox}
    \textbf{Findings 3:}
    In both two- and three-version systems, architecturally diverse combinations achieve higher $Cov\_OD$, while architecturally similar combinations achieve higher $Cer\_OD$, and this trade-off is amplified as more diverse models are added.
    By explicitly quantifying diversity and consistency, the proposed metrics provide richer insights into N-version system design than $mAP$ alone, and confirm that sufficient error coverage, as measured by $Cov\_OD$, is essential for effective model selection.
\end{findingbox}

\subsection{Relationship Between $Cov\_OD$, $Cer\_OD$, and Voting Strategies}
\label{subsec:corr-integration}
\begin{figure*}[ht]
    \centering
    \includegraphics[width=\linewidth]{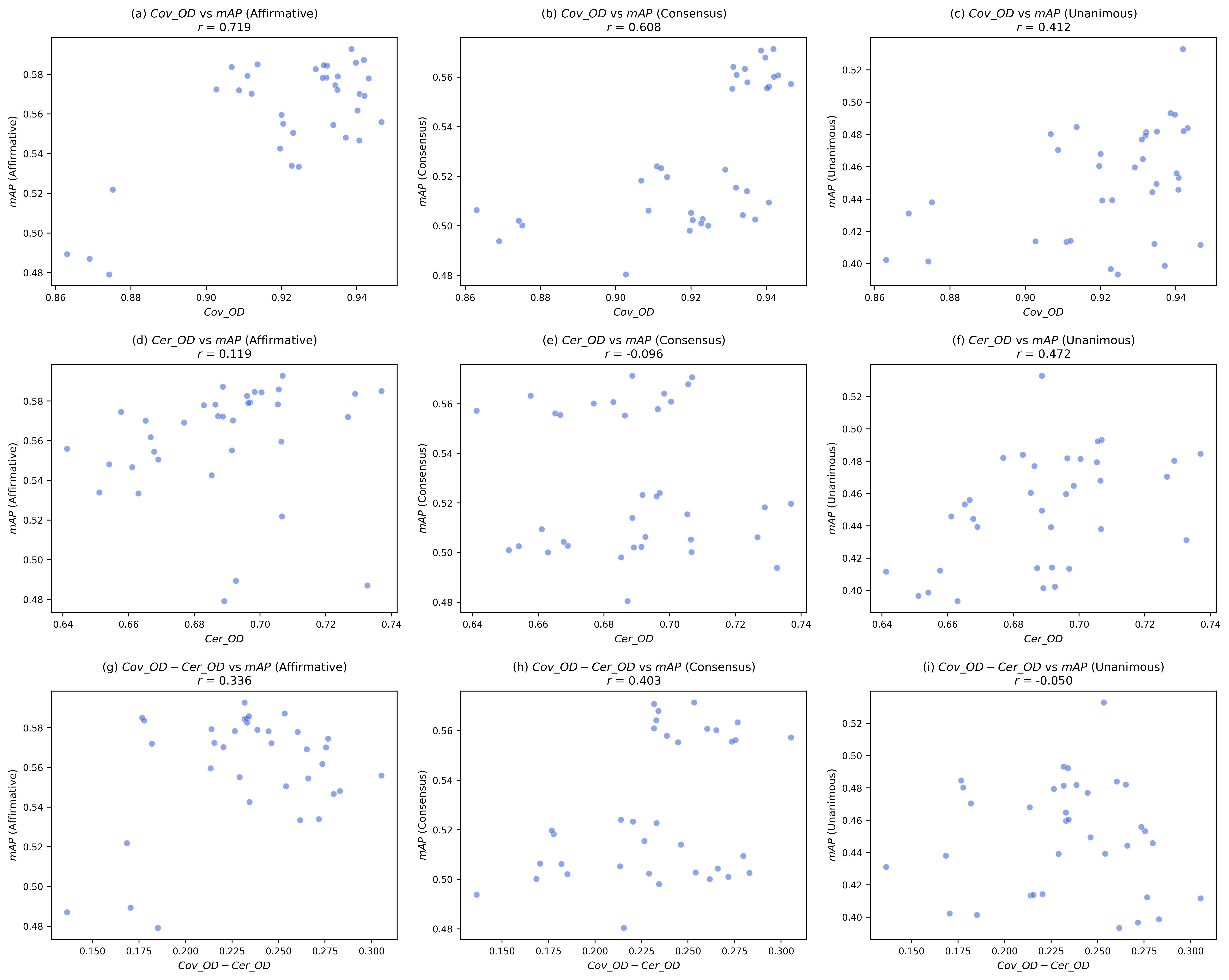}
    \caption{
    Scatter plots showing the relationships between the proposed metrics and $mAP$ across all 35 three-version configurations.
    Each point represents one configuration, and the Pearson correlation coefficient ($r$) is shown in each subplot.
    The first row (a)--(c) shows $Cov\_OD$ vs.\ $mAP$ under the affirmative, consensus, and unanimous voting strategies, respectively.
    The second row (d)--(f) shows the corresponding plots for $Cer\_OD$.
    The third row (g)--(i) shows the plots for $Cov\_OD - Cer\_OD$ (Gap), which quantifies the degree of
    complementarity between models.
    }
    \label{fig:correlation}
\end{figure*}
Any deployed N-version system must adopt a voting strategy to produce a final output, and the choice directly affects performance.
However, selecting a voting strategy based on $mAP$ creates a circular dependency, since $mAP$ requires committing to a specific voting strategy, making it impossible to use $mAP$ to inform that decision before it is made.
The proposed metrics break this dependency by characterizing the configuration prior to any voting strategy.
In this section, we investigate how $Cov\_OD$, $Cer\_OD$, and the gap can guide voting strategy selection.

Figure~\ref{fig:correlation} shows scatter plots of the relationships
between the proposed metrics and $mAP$ under each voting strategy across all three-version object detection systems. 
The Pearson correlation coefficient ($r$) represents the strength and direction of the relations.

$Cov\_OD$ exhibits a clear positive correlation with $mAP$ under the affirmative ($r = 0.719$) and consensus ($r = 0.608$) methods, since these methods retain correct detections from at least one or a majority of models, directly benefiting from higher error coverage. 
The correlation is weaker under the unanimous method ($r = 0.412$), as requiring agreement among all models limits the benefit of error coverage.

In contrast, $Cer\_OD$ shows weak or slightly negative correlations under the affirmative ($r = 0.119$) and consensus ($r = -0.098$) methods. 
High $Cer\_OD$ typically corresponds to architecturally similar models that share not only correct detections but also similar errors, providing limited complementary detections under these methods. 
Under the unanimous method, however, $Cer\_OD$ shows a moderate positive correlation ($r = 0.472$), as higher agreement in correct detections increases the likelihood that detections are retained.

The gap ($= Cov\_OD - Cer\_OD$) shows moderate positive correlations under the affirmative ($r = 0.336$) and consensus ($r = 0.403$) methods, reflecting the degree to which models complement each other's errors. 
Under the unanimous method, the gap shows no meaningful correlation ($r = -0.050$), consistent with the behavior of $Cov\_OD$.

Based on these observations, we propose the following guideline for voting strategy selection given a configuration's $Cov\_OD$ and $Cer\_OD$ values.
\begin{enumerate}
    \item Compute $Cov\_OD$, $Cer\_OD$, and the gap for the candidate configuration using per-module detection results, without committing to any voting strategy.
    \item When $Cov\_OD$ or the gap is high, coverage-oriented methods such as affirmative or consensus effectively exploit complementary detections and should be preferred. 
    When $Cer\_OD$ is high, the unanimous method is more appropriate, as consistent detections are more likely to be preserved.
\end{enumerate}

\begin{findingbox}
    \textbf{Findings 4:}
    $Cov\_OD$ and the gap correlate with mAP under affirmative and consensus voting, whereas $Cer\_OD$ correlates with unanimous voting.
    These results suggest that the proposed metrics are useful for determining the voting strategy.
\end{findingbox}

\subsection{Incremental Construction of N-Version Systems via Pairwise Analysis}
The results in Tables~\ref{tab:two_version_od_map_comparison} and~\ref{tab:three_version_od_map_comparison} reveal that when extending a two-version system by adding a third model, a configuration with high $Cov\_OD$ is more likely to be achieved by selecting a model that has low $Cer\_OD$ with both existing models, rather than simply choosing one that has high $Cov\_OD$ with either of them.

This is demonstrated concretely by starting from the best two-version pair, FCOS and SSD ($Cov\_OD$ of 0.929), and comparing the five remaining models as candidates for the third version. 
Adding RT-DETR yields the highest three-version $Cov\_OD$ of 0.947.
The pair of RT-DETR and FCOS has $Cer\_OD$ of 0.712, and the pair of RT-DETR and SSD has $Cer\_OD$ of 0.675, both of which are low.
By contrast, adding YOLOv11n reaches only $Cov\_OD$ of 0.941, despite the pair of YOLOv11n and FCOS having a slightly higher $Cov\_OD$ than the corresponding RT-DETR pair.
The difference is that the pair of YOLOv11n and FCOS has $Cer\_OD$ of 0.721, and the pair of YOLOv11n and SSD has $Cer\_OD$ of 0.715, both higher than the RT-DETR counterparts.

When extending a two-version system to a three-version system by adding a third model, the key question is how much additional coverage the third model provides beyond what the existing two models already cover. 
A high pairwise $Cov\_OD$ between the third model and each of the existing two models indicates that each pair shares few overlapping errors.
However, it does not guarantee that the third model's correct detections differ from those that the existing pair collectively provides. 
The additional coverage that the third model provides to each of the existing two models may largely overlap with the coverage that the two existing models already provide to each other. 
In such a case, the third model contributes little error coverage beyond what the existing pair collectively achieves, and the gain in three-version $Cov\_OD$ remains limited.
In contrast, when the third model has low pairwise $Cer\_OD$ with both existing models, it confirms that the third model's detection behavior differs substantially from each of them individually. 
Because its correct detections are inconsistent with those of both existing models, its errors are unlikely to coincide with the residual errors already shared by the pair. 
As a result, the third model genuinely expands the coverage beyond what the existing two models provide, leading to a larger improvement in the three-version $Cov\_OD$.
 
These findings suggest a practical strategy: when adding a model to an existing configuration, select the candidate that minimizes $Cer\_OD$ with all models already in the system.
Therefore, evaluating the uniformity of pairwise diversity at the (N-1)-version level provides a practical guideline for constructing effective N-version object detection systems. 
\begin{findingbox}
    \textbf{Findings 5:} 
    When extending a two-version system to three versions, the third model should be selected based on low $Cer\_OD$ with both existing models, not solely on high $Cov\_OD$.
    Low $Cer\_OD$ indicates that the new model's errors are unlikely to overlap with those already shared by the pair, which directly reduces the common error region and raises $Cov\_OD$.
\end{findingbox}

\section{Limitation}
\label{section-limitation}
While the proposed metrics and analyses provide insights for evaluating and designing N-version object detection systems, several limitations should be noted.

First, all experiments in this study are conducted on a custom dataset collected from the Carla simulator. 
Although this dataset provides accurate annotations, the generality of the results to other datasets and application domains is not fully validated. 

Second, the correlation analysis and observations on transitions from two-version to three-version systems are based on a limited set of model combinations. 
We exhaustively evaluated all combinations of the seven models.
Nevertheless, the number of possible combinations in practical systems is significantly larger, and the observed relationships may not fully generalize to all possible configurations.

Third, the current study focuses on only three voting strategies. 
Other voting strategies, such as Weighted Box Fusion~\cite{solovyev2021weighted}, are not considered, and the effectiveness of the proposed metrics for guiding such advanced voting strategies remains an open question.

\section{Conclusion}
\label{section-conclusion}
In this paper, we proposed new metrics, $Cov\_OD$ and $Cer\_OD$, to characterize the reliability of N-version object detection systems.
Our proposed metrics extend $Cov$ and $Cer$ to support object-level granularity, allowing for fine-grained and informative reliability analysis.
Through experiments using a dataset collected from the Carla simulator, we demonstrated that $Cov\_OD$ and $Cer\_OD$ provide more accurate and meaningful evaluations compared to previous metrics. 
We also demonstrated that the proposed metrics provide actionable design insights for N-version object detection systems. 

For future work, we plan to extend the validation to publicly available datasets such as KITTI \cite{KITTI} and COCO \cite{COCO}.
In addition, we aim to explore efficient strategies for identifying combinations of input images and object detection models that maximize the proposed reliability metrics.

\end{document}